\documentclass[10pt, conference, letterpaper]{IEEEtran}
\IEEEoverridecommandlockouts
\usepackage{amsmath,amsfonts}
\usepackage{algorithmic}
\usepackage{algorithm}
\usepackage{array}
\usepackage{textcomp}
\usepackage{stfloats}
\usepackage{url}
\usepackage{verbatim}
\usepackage{graphicx}
\usepackage{cite}

\usepackage{threeparttable}
\usepackage{times}
\usepackage{epsfig}
\usepackage{graphicx}
\usepackage{amsmath}
\usepackage{amssymb}
\usepackage{helvet} % DO NOT CHANGE THIS
\usepackage{courier}  % DO NOT CHANGE THIS
\usepackage{graphicx}  % DO NOT CHANGE THIS
\usepackage{epsfig}
\usepackage{amsmath}
\usepackage{amssymb}
\usepackage{bm}
\usepackage{subfigure}
\usepackage{booktabs}
\usepackage{multirow}
\usepackage{float}
\usepackage{bbding}
\usepackage{array}
\usepackage{color}
\usepackage{diagbox}
\usepackage{comment}
\usepackage{arydshln}
\usepackage{subfigure}
\newcolumntype{I}{!{\vrule width 1pt}}
\newlength\savedwidth
\newcommand\whline{\noalign{\global\savedwidth\arrayrulewidth
                           \global\arrayrulewidth 0.7pt}%
                  \hline
                  \noalign{\global\arrayrulewidth\savedwidth}}
\newlength\savewidth

\begin{document}

\title{Joint Channel Estimation and Feedback with Masked Token Transformers in Massive MIMO Systems}

%\author{Mingming Zhao*, Lin Liu*, Lifu Liu, Mengke Li\corref{cor1}, Qi Tian}
        % <-this % stops a space
\author{Mingming Zhao*, Lin Liu*, Lifu Liu, Mengke Li\dag, Qi Tian \thanks{*Mingming Zhao and Lin Liu contribute equally to this artile and should be considered co-first authors.}\thanks{\dag Mengke Li  is corresponding author.}
  }

%Mingming Zhao*, Lin Liu*, Lifu Liu, Mengke Li\corref{cor1}, Qi Tian

%}

% \thanks{*The authors contribute equally to this artile.}
% \thanks{Corresponding author: Mengke Li}

\markboth{Journal of \LaTeX\ Class Files,~Vol.~14, No.~8, August~2021}%
{Shell \MakeLowercase{\textit{et al.}}: A Sample Article Using IEEEtran.cls for IEEE Journals}


\maketitle

\begin{abstract}
The downlink channel state information (CSI) estimation and low overhead acquisition are the major challenges for massive MIMO systems in frequency division duplex to enable high MIMO gain.
Recently, numerous studies have been conducted to harness the power of deep neural networks for better channel estimation and feedback. 
However, existing methods have yet to fully exploit the intrinsic correlation features present in CSI. 
As a consequence, distinct network structures are utilized for handling these two tasks separately.
To achieve joint channel estimation and feedback, this paper proposes an encoder-decoder based network that unveils the intrinsic frequency-domain correlation within the CSI matrix.
%The proposed method comprises two main components. 
The entire encoder-decoder network is utilized for channel compression.
To effectively capture and restructure correlation features, a self-mask-attention coding is proposed, complemented by an active masking strategy designed to improve efficiency. 
The channel estimation is achieved through the decoder part, wherein a lightweight multilayer perceptron denoising module is utilized for further accurate estimation.
Extensive experiments demonstrate that our method not only outperforms state-of-the-art channel estimation and feedback techniques in joint tasks but also achieves beneficial performance in individual tasks.
\end{abstract}


\section{Introduction}\label{sec1}
Massive multiple-input multiple-output (MIMO) technology is a crucial component of 5G-Advanced and 6G networks. 
It increases the number of transmission antennas and improves cell transmission quality and cell system capacity~\cite{MIMO2015introduction, TariqF20206G}. 
%当前系统损失的原因与本文优化目标相关，优化目标在系统里面的作用
The utilization of large arrays of analog antennas in MIMO systems introduces transmission overhead and noise interference, leading to notable performance degradation, particularly in frequency division duplex (FDD) systems.
There is no reciprocity~\cite{RusekF2012ScalingupMIMO} in FDD systems, and as a result, the base station requires the user to transmit back the downlink (DL) channel state information (CSI) to build the communication link to obtain the MIMO gain. 
On the one hand, the size of DL CSI matrix demands a significant overhead but the number of available bandwidths remains limited, thus, channel compression feedback is necessary. 
On the other hand, achieving an accurate feedback channel requires pilot channel estimation.

%当前信道估计和压缩的基础问题
%解决这些问题以前研究的sota
Previous studies have explored various kinds of methods for optimizing channel estimation and/or feedback to improve the accuracy of the downlink channel. 
For the channel estimation, the least squares (LS)~\cite{SongWG2006ChannelEst} and minimum mean square error (MMSE)~\cite{MaJF2014SDataAidChannelEst} are two commonly used classical methods, but LS estimations suffer from significant noise, and applying the MMSE method in practice poses challenges due to the
lack of access to accurate channel information beforehand.
Recently, the learning-based methods include CNN-based ~\cite{MehranSoltani2019DeepLC, YuJin2020ChannelEF} and attention-based~\cite{DianxinLuan2022AttentionBN} networks can effectively increase the estimation accuracy.
For channel compression, the 3GPP release-16 (Rel-16) standards specify the utilization of Type I and improved Type II (eType II) codebooks for the CSI feedback system ~\cite{3GPP2020release16}. 
The encoder-decoder based structure~\cite{ChaoKaiWen2018CsiNet, LuZ2021CRNet, CaoZ2021ConvCsiNet,SunY2020AnciNet,GuoJ2020CsiNet+} can further improve the CSI feedback performance.
For joint channel estimation and feedback, deep-learning-based methods ~\cite{MaX2020JointE2E, ChenTong2020DeepLF} achieve significant improvement. 
Attention-based methods are the most well-known deep-learning-based methods, which improve the performance of channel estimation and feedback by strengthening the connection between different channels~\cite{NIPS2017_3f5ee243, XiaoH2021EVCsiNet-T, YangXu2021TransformerEC, JiabaoGao2021AnAD}.
However, previous methods have been limited to performing either a single task or joint tasks with two independent networks or methods.

%\textcolor{red}{TODO-1}
%However, these DL-based methods may cause channel estimation errors when the number of given pilots is small or unsatisfactory compression performance when the bit number is low. 
%\textcolor{red}{TODO-2}
%Our proposed masked token scheme can learn a better bijection between deep features and channel information by adding learnable tokens.
To explicitly incorporate the channel estimation and feedback into one unified structure, we propose an attention-based method named \textbf{FlowMat}. 
It is based on encoder-decoder architecture and makes full use of the characteristics of CSI and each part of the network.
Based on the observation that inherent frequency-domain correlations exist among channel matrices, it becomes possible to achieve mutual substitution of frequency domain data.
We leverage this characteristic of the channel matrix and propose self-attention coding to facilitate the acquisition of relevant features.
Then, channel compression can be achieved through extract highly correlated features. 
We proposed an active mask technique to acquire these highly correlated features.
The learnable mask tokens make the encoder learn a better mapping between deep features and channel information. 
Then, channel completion and recovery are achieved by decoding mask tokens.
For channel estimation, we use the decoder part. 
In order to reduce environmental noise and complement the $H$ channel, we introduce a lightweight multilayer perceptron denoising module.
The proposed FlowMat framework enables estimation and feedback with high performance and low computational overhead, which outperforms state-of-the-art approaches.
Our main contributions are summarized as follows:
\begin{itemize}
\item We reveal the inherent frequency-domain correlation that existed among channel matrices.
To leverage this correlation effectively, we devise the frequency domain vector as the fundamental channel feature base unit. 
This approach bears a resemblance to the concept of employing non-overlapping patches in Computer Vision (CV) and word tokens in Nature Language Processing (NLP).
%We reveal the intrinsic frequency-domain correlation across CSI matrix, and design the frequency domain vector as channel feature base unit, which is similar to the CV non-overlapping patch and NLP word token.
\item We propose an innovative token transformer named FlowMat, which can achieve simultaneous channel estimation and feedback in the downlink scenario. 
FlowMat is uniquely equipped to harness the frequency domain's inherent correlation to achieve both completion and compression tasks efficiently. 
%We propose a novel masked token transformer solution called FlowMat for joint channel estimation and feedback in the downlink scenario. The FlowMat can take advantage of the correlation between frequency domain to achieve completion and compression, by using masking and interpolation.
\item Extensive experiments on joint channel estimation and feedback tasks show that our proposed method outperforms state-of-the-art approaches.
Moreover, applying the proposed method to individual channel estimation or feedback can achieve performance gain as well.
\end{itemize}

% The other sections are arranged as follows:
% %
% Section II provides an overview of several related methods for channel estimation and channel feedback, including ChannelNet, attention networks, and EvcsiNet. Section III presents our proposed FlowMat method for joint channel estimation and channel feedback. Section IV presents simulation results comparing our method with different existing methods. Finally, Section VI summarizes the key contributions of this paper.

\section{Related Work}
\label{sec2}
\subsection{Channel Estimation}
%For channel estimation, there have been recent studies on the use of neural networks.
Deep learning has recently become the main tool of numerous studies as a promising approach for channel estimation.
These studies can be categorized into two types: non-attention-based and attention-based methods.
%such as ChannelNet, AttentionNet, CBDNet and Attention-Aid.
For no-attention-based methods, ChannelNet~\cite{MehranSoltani2019DeepLC} is the first deep-learning-based work to solve the channel estimation problem.
It utilizes a super-resolution network SRCNN~\cite{dong2015image}, and a denoising network DnCNN for channel completion and denoising, respectively.
Chun et al.~\cite{chun2019} propose a two-phase model which estimates channels in time domain. 
The first phase employs a pilot-aid model consisting of a two-layer MLP and a CNN, while the second phase utilizes a data-aid model.
Jin et al.~\cite{YuJin2020ChannelEF} exploit an image denoising network called CBDNet for channel estimation. The network is composed of two main sub-networks: a noise level estimation sub-network and a non-blind denoising sub-network.
He et al.~\cite{HengtaoHe2018DeepLC} utilize a learned denoising-based approximate message-passing network for beamspace millimeter-wave (mmWave) MIMO systems.
For attention-based methods, AttentionNet~\cite{DianxinLuan2022AttentionBN} utilizes transformer blocks for channel attention.
Gao et al.~\cite{JiabaoGao2021AnAD} and Mashhadi et al.~\cite{Mahdi2021Pruning} propose the CNN channel attention module and CNN non-local blocks for channel estimation, respectively.
%
% Our proposed FlowMat also exploits attention-based methods. 
% It needs to be noted that the FlowMat can learn better mapping between deep features and channel information by adding learnable masked tokens and self-mask-attention, which enables efficient channel recovery with high performance. 
% Due to the presence of noise during pilot transmission, we introduce the MLP-Mixer\cite{NIPS2021_MLPMIXER} as a lightweight encoder to perform denoising

\subsection{Channel Feedback}
In addition to channel estimation, channel feedback is also a crucial technology in the physical layer of MIMO systems.
Wen et al.~\cite{ChaoKaiWen2018CsiNet} propose CsiNet which is based on a CNN autoencoder structure.
%Its main network is a convolutional neural network (CNN). 
In CsiNet, the encoder is designed for CSI compression, while the decoder is responsible for recovering both the user equipment (UE) and base station (BS).
Following, inspired by CsiNet, a series of studies~\cite{LuZ2021CRNet, CaoZ2021ConvCsiNet,SunY2020AnciNet,GuoJ2020CsiNet+} design various kinds of CNNs to further improve the CSI feedback performance. 
Mashhadi et al.~\cite{MashhadiMB2020DeepCMC} propose an image compression approach, which incorporates rate-distortion cost and arithmetic entropy coding to achieve the minimum bit overhead.
The long short-term memory (LSTM) networks~\cite{LuC2018RNN,Chen2021ImCSI} are introduced into the encoder and decoder to make full use of the correlation extracted from subcarriers.
The majority of  CNN and LSTM-based methods concentrated on full channel state information (F-CSI) feedback, without considering the character of channels, such as correlation, which can lead to unrepresentative compressed features and low recovery accuracy.
% The whole downlink channel matrix is compressed and subsequently recovered at the encoder and decoder, respectively.
The current 5G system primarily uses CSI feedback, which is based on the compression and feedback of the eigenvector of the channel matrix, as stated in 3GPP. 
Liu et al.~\cite{LiuW2021EVCsiNet} propose EVCsiNet, which is a CNN-based structure that utilizes eigenvector features. 
In recent times, attention-based techniques have garnered significant popularity in the domains of CV and NLP.
Xiao et al.~\cite{XiaoH2021EVCsiNet-T} propose EVCsiNet-T for channel feedback, which compresses the channel eigenvector and the attention mechanism for channel recovery.

% The proposed FlowMat is the masked token transformer. We have developed a novel active masking mechanism for dropping tokens and embedding masked tokens, using the self-mask-attention achieves encoding and decoding, which enables efficient compression and recovery with high performance and low computational overhead.

\subsection{Joint Channel Estimation and Feedback}
All of the deep-learning-based CSI feedback schemes mentioned above assume that the downlink channels are accurately known to the user, which is not feasible in practical communication systems. 
Obtaining precise channel information is crucial for base stations. 
Therefore, recent investigations have explored the application of neural networks for joint channel estimation and feedback optimization.
Ma et al.~\cite{MaX2020JointE2E} propose a deep neural network (DNN) architecture that jointly designs the pilot signals and channel estimation module end-to-end to avoid performance loss caused by separate designs. 
Chen Tong et al.~\cite{ChenTong2020DeepLF} propose JCEF, where two networks are constructed to perform explicit and implicit channel estimation and feedback, respectively. 
% The proposed FlowMat is the whole framework for channel estimation and compression. 
% For the channel compression, we use the whole FlowMat including the encoder-decoder framework. 
% For the channel estimation, we mainly use FlowMat's decoder to achieve the completion and a new adding denoise network as encoder. 
% So, the FlowMat framework enables efficient estimation and compression with high performance and low computational overhead.%and outperforms state-of-the-art approaches.

\section{Preliminaries}\label{sec3}

\subsection{System Model}

\subsubsection{Massive MIMO system}
For a typical massive MIMO system in FDD mode, we consider that the system equips with $N_t(\geq1)$ transmit antennas at the BS and $N_r(\geq1)$ receive antennas at UE side. Orthogonal frequency division multiplexing (OFDM) with $N_c(\geq1)$ subcarriers is adopted consisting of 4 resource blocks (RBs). In the downlink phase, the corresponding the $i$-th received signal component ${y_i}\in{\mathbb{C}^{N_c\times{1}}}$ can be denoted as, 

\begin{equation}
\label{eq:H_system}
y_i = H_iP_is_i + n_i,
\end{equation}
where ${s_i}\in{\mathbb{C}^{N_c\times{1}}}$ is the matrix of the transmitting signal to the $i$-th receive antenna. 
$H_i\in{\mathbb{C}^{N_c\times{N_t}}}$ represents the channel frequency response vector, which is estimated by the pilot-based channel estimation. 
$P_i\in{\mathbb{C}^{N_t\times{N_c}}}$ is the corresponding precoding matrix.%, which beamforming and eliminating user interference in the BS. 
$n_i\in{\mathbb{C}^{N_c\times{1}}}$ represents the additive noise. 
Then, the channel matrix $H_i$ in the frequency-spatial domain can be obtained by stacking $h_{i,j}$ in the frequency domain as follows: $H_i=[h_{i,1}, h_{i,2},...,h_{i,k}], h_{i,k}\in{\mathbb{C}^{N_t\times{1}}}$. 
However, this matrix size is unacceptably large for direct feedback in a massive MIMO system. 

%Previous works (e.g., EZF~\cite{jindal2006mimo}) have also made efforts to exploit eigenvector decomposition and precoding to achieve lightweight precoding and establish efficient communication links.
The potential of eigenvector decomposition (e.g., EZF~\cite{jindal2006mimo}) can be exploited to achieve lightweight precoding and establish efficient communication links.
After channel estimation at UE, the corresponding eigenvector matric $w\in{\mathbb{C}^{N_t\times{N_c}}}$ is composed of $w_{k}\in{\mathbb{C}^{N_t\times{1}}}$ with normalization  $\Vert{w_{k}}\Vert^2=1$. 
$w_{k}$ is the $k$-th subcarrier of CSI.
$W$ can be utilized as the downlink precoding vector and calculated using eigenvector decomposition.
The based eigenvector decomposition and precoding system model can be calculated as:
\begin{equation}
H_k^HH_kw_k = \lambda_{k}{w_k},
\end{equation}
\begin{equation}
\label{eq:W_system}
y_i = w P_is_i + n_i,
\end{equation}
where $\lambda_i$ represents the maximum eigenvalue of $H_k^HH_k, H_k\in{\mathbb{C}^{N_r\times{N_t}}}$, and $()^{\mathrm{H}}$ is Hadamard transpose.
Therefore, to build an accurate communication link and obtain MIMO gain, pilot-based channel estimation and channel feedback are key challenges in FDD system.
\subsubsection{Channel estimation}
Before building the complete communication link, the BS needs to transmit the pilots to the receiver, because downlink channels are  unknown to the transmitter. The pilot symbol-based channel estimation often performs quite well for tracking the sudden change of the channel, especially the fading channels in $N_p (1\leq{N_p}\leq{N_c})$ equi-spaced pilot subcarriers. However, the size of the density and spatial link noise directly affect the accuracy of channel recovery. 
Conventional channel estimation methods, such as the least-squares (LS) technique are widely used by minimizing the mean squared error (MSE) between received $y_{i,p}, y_{i,p}\in{\mathbb{C}^{N_p\times{N_t}}}$, to give an estimate of $\hat{H}_{i,p}$, the frequency domain LS estimation is given by
\begin{equation}
\label{eq:LS Model}
\hat{H}_{i,p}=\mathop{argmin}_{\hat{H}_{i,p}} {\Vert{y_{i,p}-\hat{H}_{i,p}\odot{s_{i,p}}\Vert}},
\end{equation}
where $\hat{H}_{i,p}, s_{i,p}\in{\mathbb{C}^{N_p\times{N_t}}}$ denotes the estimated and transmitted pilot signals respectively and $\odot$ denotes the Hadamard product. %element-wise mathematical multiplication. 
Then, interpolation will be conducted to obtain the channel responses at other subcarriers to obtain the whole $H_i$.
The LS technique is known for its ease of implementation and extremely low complexity. 
% We consider using the high-density and low-density symbols as pilots in our article. 
\subsubsection{Channel feedback}
In FDD systems, the UE will send the estimated downlink CSI back to the BS after receiving the pilot symbols and estimating the downlink channel matrix. Then, the BS may create relevant precoding vectors to reduce user interference and enhance communication quality.
In massive MIMO systems, the CSI matrix $H\in{\mathbb{C}^{N_c\times{N_t}\times{N_r}}}$, which has a total of $2\times{N_c}\times{N_t}\times{N_r}$ real number components, will result in significant feedback overhead. The compressibility of the CSI matrix has been extensively studied in literature since it is desirable in actual systems to minimize the feedback parameters.
Adopting the multi-stream downlink transmission at the UE side, the corresponding eigenvector for the subcarriers, the BS can be directly used as the downlink precoding vector. Using the eigenvector as channel matrix can also reduce the $N_r$ dimensions of the receiving antenna. So the size of the channel matrix is $2\times{N_c}\times{N_t}$ for the eigenvector. To further decrease the feedback overhead of DL CSI and enable accurate CSI recovery at the BS, we apply the typical DNN-based method for CSI compression and reconstruction.

\subsection{Key Observation for Wireless Channel}
We conducted a comprehensive analysis of wireless channel and eigenvector characteristics, which motivates us to pursue carefully crafted network designs.

% \begin{figure}[t]
%  		\centering
%  		\includegraphics[width=1.\linewidth]{MinG_Fig/ChannelFeature.png}
%  		\caption{Illustration of correlation features in the spatial frequency.}
%  		\label{fig:ChannelFeature}
%  \end{figure}

\begin{figure}[t]
   \centering   
   \subfigure[$H$ Matrix]{\includegraphics[width=0.495\linewidth,height=0.445\linewidth]{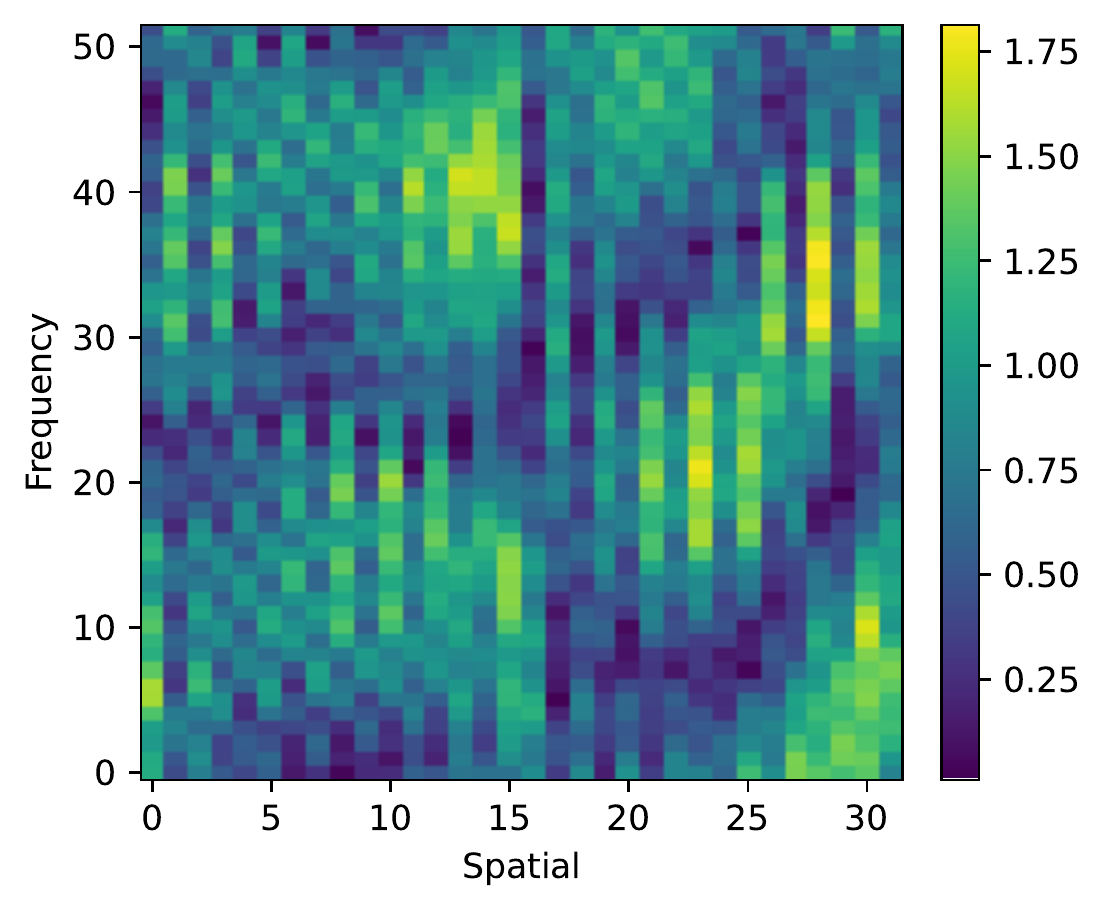}\label{fig:CSI_1}}
   \subfigure[$H$ correlation Matrix]{\includegraphics[width=0.48\linewidth,height=0.45\linewidth]{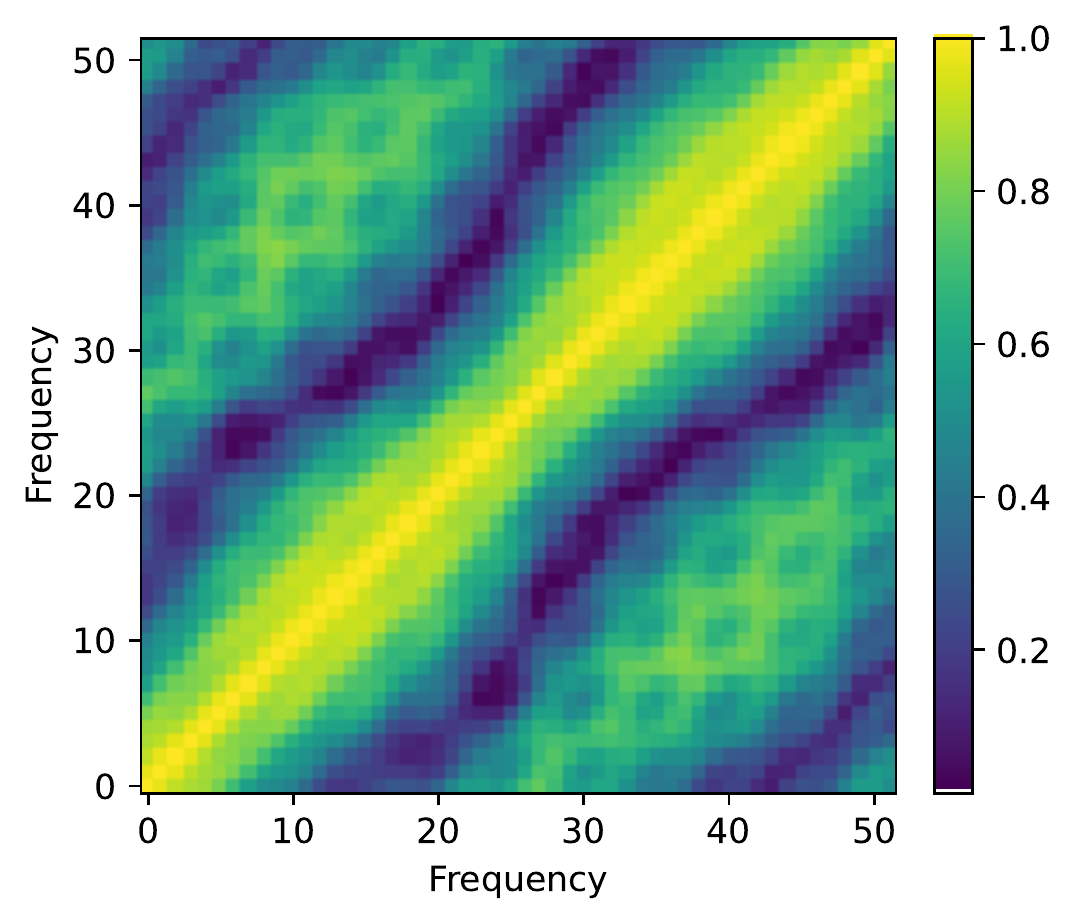}\label{fig:CSI_3}}
   \\
   \subfigure[$w$ Matrix]{\includegraphics[width=0.495\linewidth,height=0.445\linewidth]{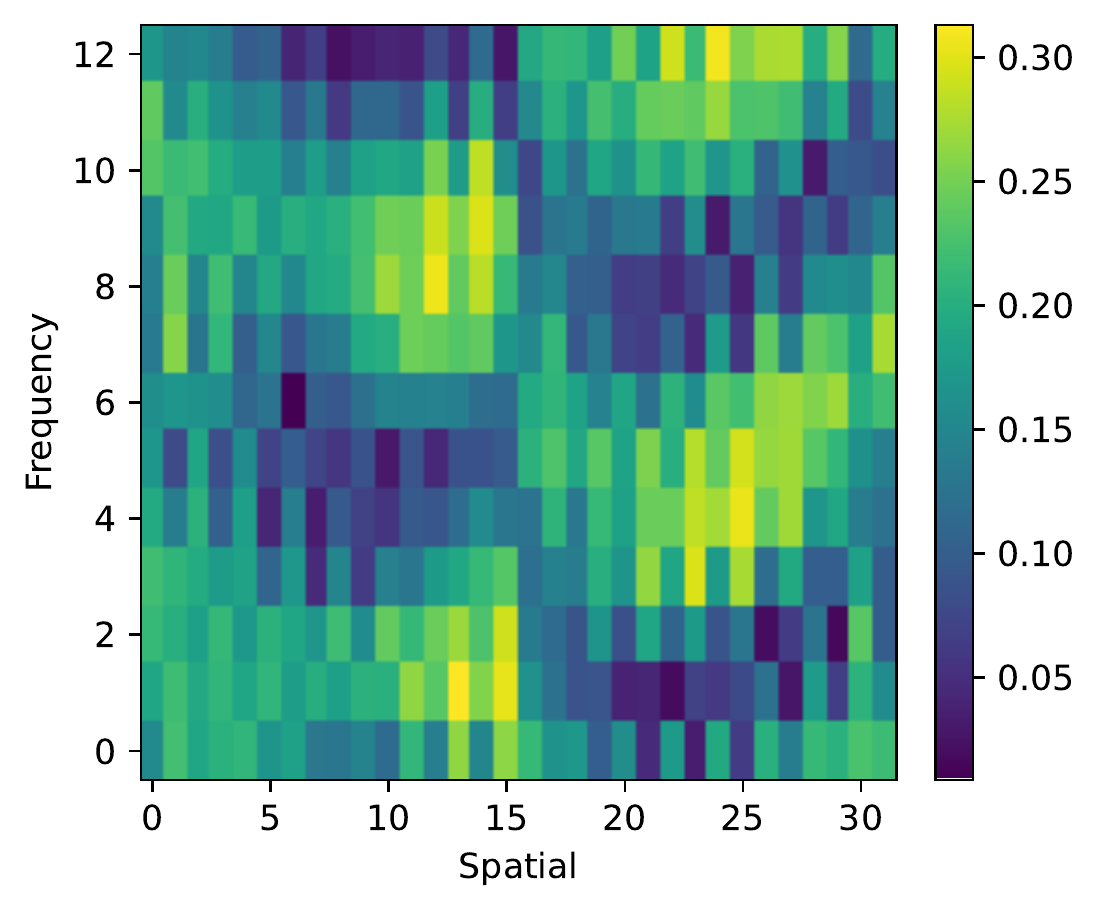}\label{fig:CSI_2}} 
   \subfigure[$w$ correlation Matrix]{\includegraphics[width=0.48\linewidth,height=0.45\linewidth]{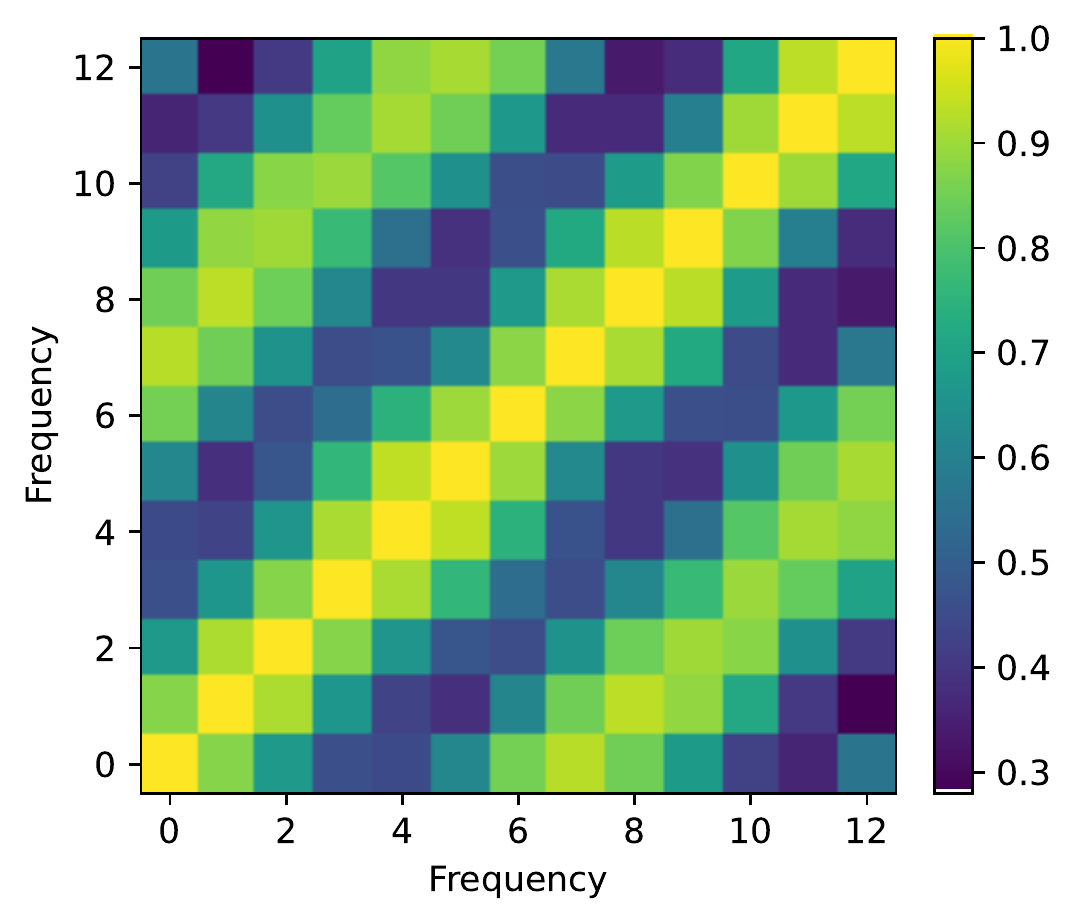}\label{fig:CSI_4}}
\caption{Illustration of correlation features in the spatial-frequency domain.}
\vspace{-6pt}
\label{fig:ChannelFeature}
\end{figure}
 
The base station builds communication links by recovering the DL CSI matrix. 
We mainly analyse the measured DL CSI matrix $H\in{\mathbb{C}^{N_c\times{N_t}\times{N_r}}}$ without noise and eigenvector $w$ in the spatial-frequency domain. 
We perform frequency-domain correlation calculations on $H_i, i=1$ and $w$ separately. 
The results are shown in Fig~\ref{fig:ChannelFeature}.
Fig.~\ref{fig:CSI_1} and Fig.~\ref{fig:CSI_3} have frequency domain similarity. 
In Fig.~\ref{fig:CSI_2} and Fig.~\ref{fig:CSI_4}, the similarity near the diagonal of the matrix is significantly higher, and it is a wealth of local information. 
The correlations are observed to be high even at far distances. 
The DL CSI has a certain level of global information. 
We thus find the frequency domain as base unit and the spatial as vector data. 
The feature token vector is similar to CV patch and NLP word token unit. 
Therefore, we utilize the frequency token unit and the self-correlation underlying characteristic to build the joint model for channel compression and channel estimation.
 
 % self-supervised learning to achieve the compression by masking part feature token vectors and completion by adding masked feature token vectors for channel compression and channel estimation.

\section{SYSTEM DESIGN}
\subsection{Proposed Methodology: FlowMat}

\begin{figure}[t]
 \centering
 \includegraphics[width=0.75\linewidth]{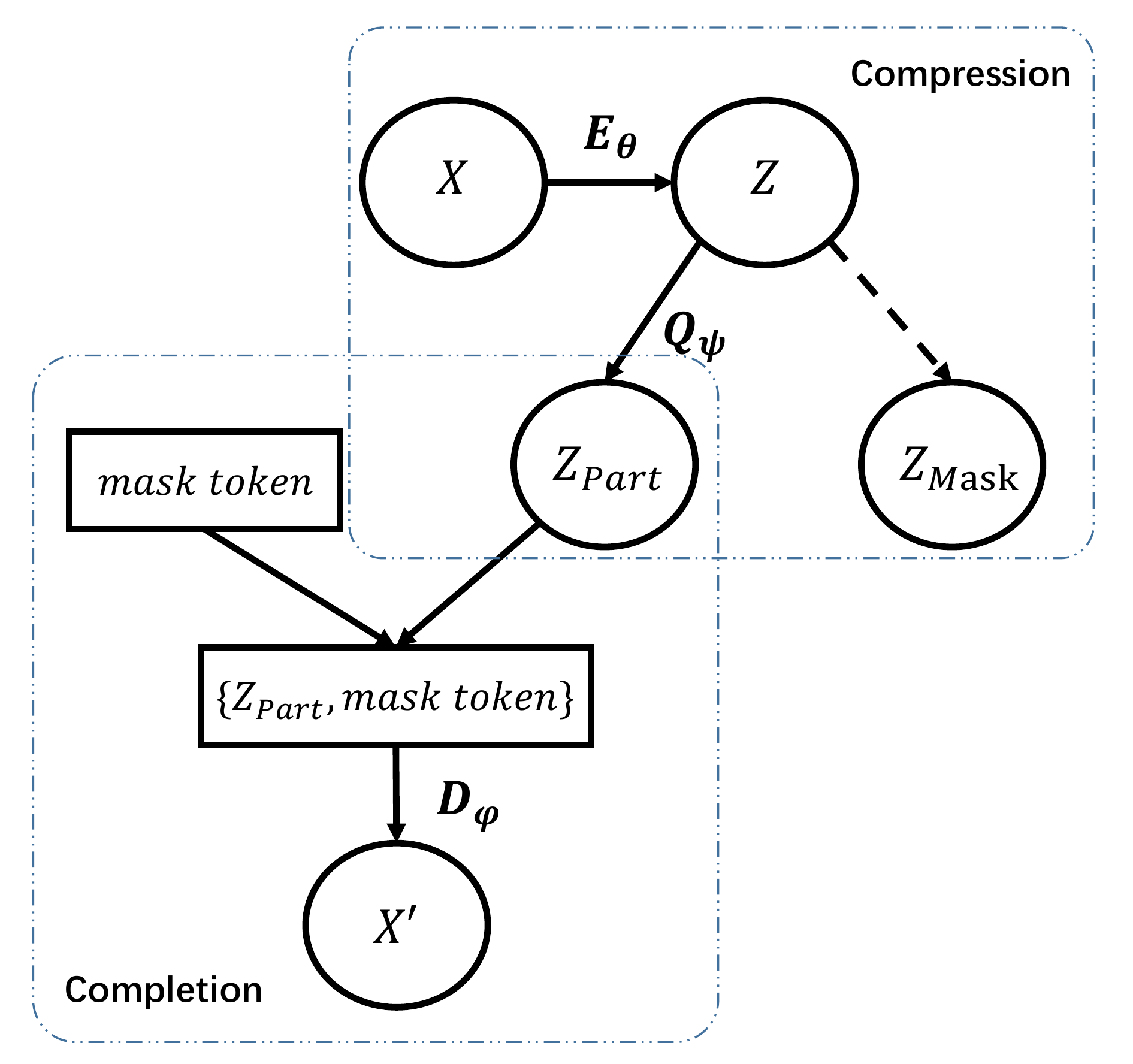}
 \caption{The compressing and completing processes of FlowMat. In a hierarchical data-encoding process, the latent variables (e.g., $Z$) represent high-level information. By active masking, $Z_{Part}$ is the compact representation, which recovers the origin data with embedding mask token and decoding.}
 \label{fig:Backbone_TheoryFlow}
 \vspace{-6pt}
 \end{figure}

Our idea of the masked token is inspirited by the work Masked AutoEncoders (MAE) from He et al.~\cite{MaskedAutoencoders2021}, where a self-supervised learning method in CV is proposed.
They divide an image into non-overlapping patches and mask random patches of them and the encoder only processes the unmasked and visible patches.
The decoder reconstructs the original image from deep representations of the encoder and masked tokens.
Each mask token is a shared learning vector indicating the existence of missing patches to be restored.
The loss function is the mean squared error (MSE) between the reconstructed images and the original images.
During pre-training, it masks the input image in a large proportion (about 75\%) and then learns to restore the complete image.
This method allows the network to learn a large-capacity model with good generalization.
The pre-trained encoder can be applied to downstream tasks such as classification and segmentation and achieve better performance than many other pre-training pipelines (such as MoCo v3~\cite{XinleiChen2021AnES} and BEiT~\cite{HangboBao2021BEiTBP}). 

\subsubsection{Problem Definition}
% There has been some work recently trying to understand MAE's theory~\cite{Cao_Xu_Clifton,CVPR2023_UnderstandingMAE}, in which~\cite{CVPR2023_UnderstandingMAE} considers the data-generating process as a hierarchical latent variable model. However, these works mainly prove that the completion of data features can be achieved by embedding $mask$ $token$ after random mask of the original data. 
% With reference to these theories, 
we propose that active masking and self-mask-attention coding method $FlowMat$ can be used to achieve compression and completion, which have been fully verified in channel estimation and feedback reconstruction.
We use a sequence $X=\{x_1,...,x_N\}, x_n\in{C^{1\times N_t}}$ to uniformly express the channel matrix $H$ and the eigenvector matrix $V$. 
We divide the data in both the frequency domain and spatial domain into two-dimensional information. 
The frequency domain vector $x_n$ serves as the channel basic unit, which can be analogously considered as a token in the fields of CV and NLP.
%The channel basic unit is also called token similar to the CV and NLP.
%The channel basic unit is known as a token, which is similar to those used in computer vision (CV) and natural language processing (NLP).
%
%The traditional DNN compression is mainly to flatten all the encoder data, and then realize it through DNN down-sampling, and reconstructing the compressed data using the up-sampling.
Traditional DNN compression techniques primarily involve flattening all the encoder data, downsampling it through DNN, and then reconstructing the compressed data using upsampling methods.
However, such an operation obviously does not make full use of the characteristics of the wireless channel.
In order to make full use of the correlation information found between the frequency domains, we actively mask a part of the tokens to achieve compression, then insert the $mask$ $token$ into the compressed tokens' subset as a complete matrix to decode and recover the whole original channel matrix.
 
\subsubsection{Model Design}
The encoder-decoder framework for compression and completion is formulated as:
%We apply the typical DNN-based encoder-decoder framework for compression and completion.
% Our compression way is selecting a subset latent variable after the encoder, and the other tokens will be masked and dropped.

\begin{equation}
Z = E_\theta(X),
\end{equation}
\begin{equation}
Z_{Part}, Z_{Mask} = Q_{\psi}(Z),
\end{equation}
where $X$, $Z$, $Z_{Part}$, and $Z_{Mask}$ are input data, compressed data, learnable useful data, and mask data, respectively.
$E_\theta$ and $Q_{\psi}$ denote encoder and query. 
To reconstruct and restore the latent variable $Z_{Part}$, we first substitute the upsampling technique with the incorporation of $mask$ $token$.
Subsequently, this modified latent variable is transmitted to the decoder.
This process is mathematically expressed as:
\begin{equation}
X' = D_\phi(Z_{Part}, mask token).
\end{equation}
The entire compressing and completing processes are illustrated in Fig.~\ref{fig:Backbone_TheoryFlow}. 
The objective of our study is to identify a set of encoding, query, and decoding functions, their parameters are denoted as $\theta$, $\phi$ and $\psi$, respectively.
This parameter set can be obtained by minimizing the discrepancy between the original matrix $X$ and its reconstructed counterpart $X'$:
\begin{equation}
(\theta, \phi, \psi) = \mathop{argmin}_{\theta, \phi, \psi} {\Vert{X-X'}\Vert}.
\end{equation}

\subsubsection{Query, Encoder and Decoder in FlowMat} 
For effective compression and completion through subset acquisition and mask token insertion, meticulous design of querying strategies, the encoder, and decoder is essential.
This section introduces these three components in detail.
%In the ensuing discussion, we will expound upon these two components.

\textbf{Query}: According to the number of compressed target frequency domain units, We generate random real vectors $query$, $query\in{R^N}$. The $query$ is a vector that can be updated from learning. Each time the target subset is selected as $Z_{Part}$, the set of $topK$ positions of the query will be selected according to the compression rate to achieve compression. 

\begin{figure}[t]
\centering
\includegraphics[width=0.35\textwidth,height=0.6\linewidth]{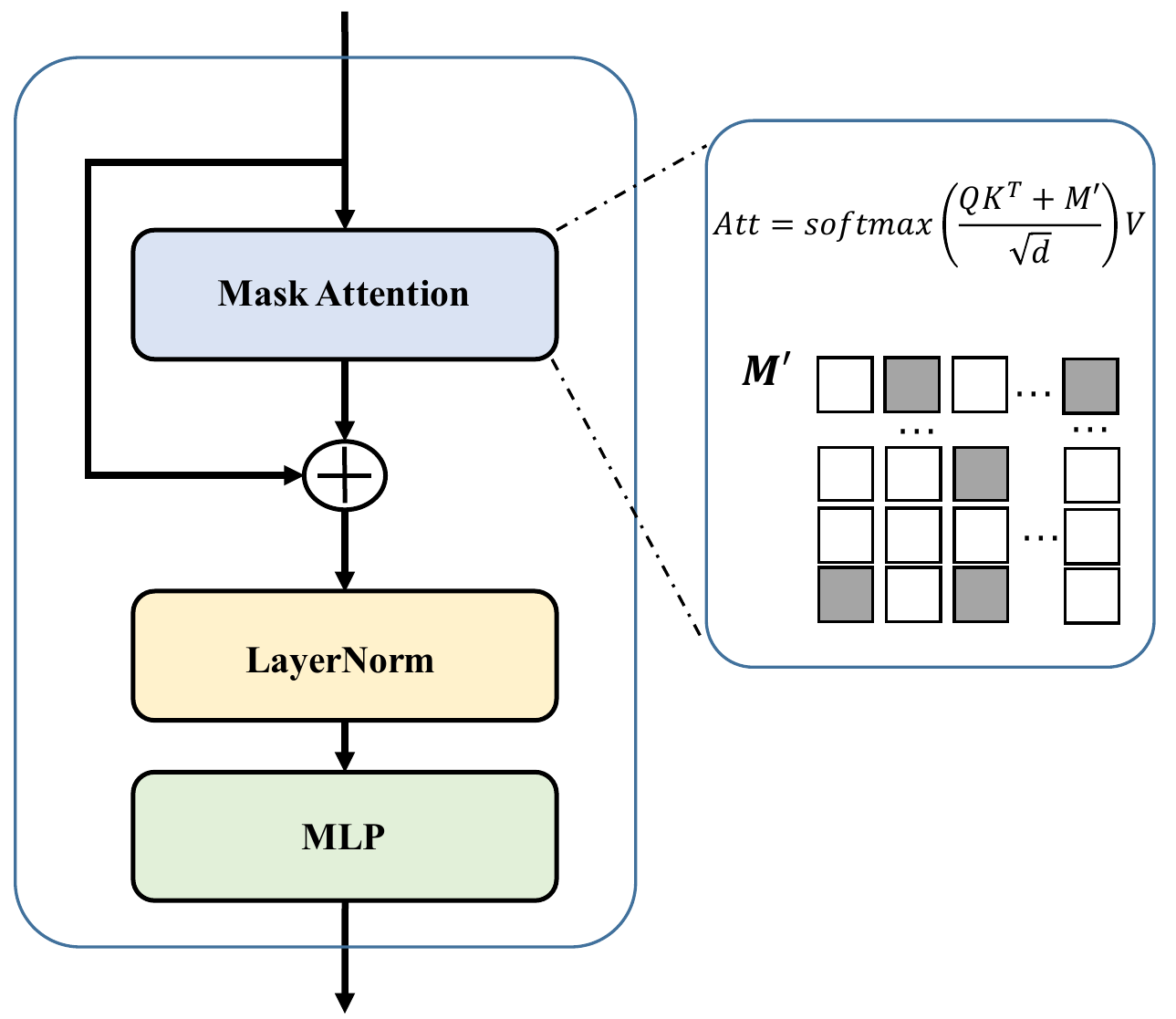}
\caption{The structure of a single mask-attention stage in MAT. `$\textbf{\textit{M}}'$' represents the proposed mask attention matrix. }
\label{fig:MaskAttention}
\vspace{-6pt}
\end{figure}

\textbf{Encoder}: We achieve compression by selecting a subset of $X$. 
To obtain high-quality reconstruction, $Z$ should be compact and representative enough.
%sufficient information about the target subset, so that we can better decode the masked data implement refactoring. 
We leverage the self-attention mechanism within the transformer structure to obtain fully correlated latent variables.
Moreover, we introduce a specialized transmission path mask $M'$ at the last block to facilitate the coding of global information to the target set.
%The encoder is calculated by:
The calculation of the encoder is:
\begin{equation}
\begin{aligned}
Y_1 &= XW_1 + Pos, \\
[Q, K, V] &= [Y_1W_Q, Y_1W_K, Y_1W_V], \\
Y_2 &= Softmax(\frac{QK^T + M'}{\sqrt{d}})V, \\
Z &= W_2Y_2,
\label{eq:encoder}
\end{aligned}
\end{equation}
where $Pos$ is the position embedding. $W_{*}, (*={1,2,Q,K,V})$ are learnble parameters. In Enq.~\ref{eq:encoder}, the mask $M'$ is:
\begin{equation}
M_{r,h}'= 
\left\{
\begin{aligned}
      & 0, \text{if token $h$ is masking} \\
      & 1, \text{if token $h$ is non-masking}
\end{aligned},
\right.
\end{equation}
where $r, h \leq{N_c}$ represents the number of the subcarriers. Fig.~\ref{fig:MaskAttention} illustrates this single mask-attention stage.

\textbf{Decoder}: 
% \textcolor{red}{The goal of reconstruction is a subset of the hidden variables of the original matrix.}
%
We initialize the completion matrix by inserting mask tokens. 
The decoder also uses the self-attention mechanism to pass effective information in $Z_{Part}$ to the mask token.%
%and realizes data reconstruction through multi-layer message passing. 
Data reconstruction is realized through multi-layer message passing.
We further add a transfer path into the first layer to embed the mask map. %to improve the transfer of valid information to the mask token, and avoid the pollution of valid information by initial invalid information. 
We introduce a transfer path into the first layer specifically for embedding the inverse mask $M'$.
This transfer path enhances the transferability of relevant information to the mask token, meanwhile preventing the contamination of valid information by initial invalid data.
%The decoder way is calculated as:
The process of decoder is formulated as:
\begin{equation}
\begin{aligned}
Y_3 &= [Z_{Part}, Mask\_Token], \\
Y_4 &= Y_3W_3 + Pos, \\
[Q, K, V] &= [Y_4W_Q, Y_4W_K, Y_4W_V], \\
Y_5 &= softmax(\frac{QK^T - M'}{\sqrt{d}})V, \\
X' &= W_1Y_5.
\end{aligned}
\end{equation}

\begin{figure*}[t]
\centering
\includegraphics[width=0.90\textwidth]{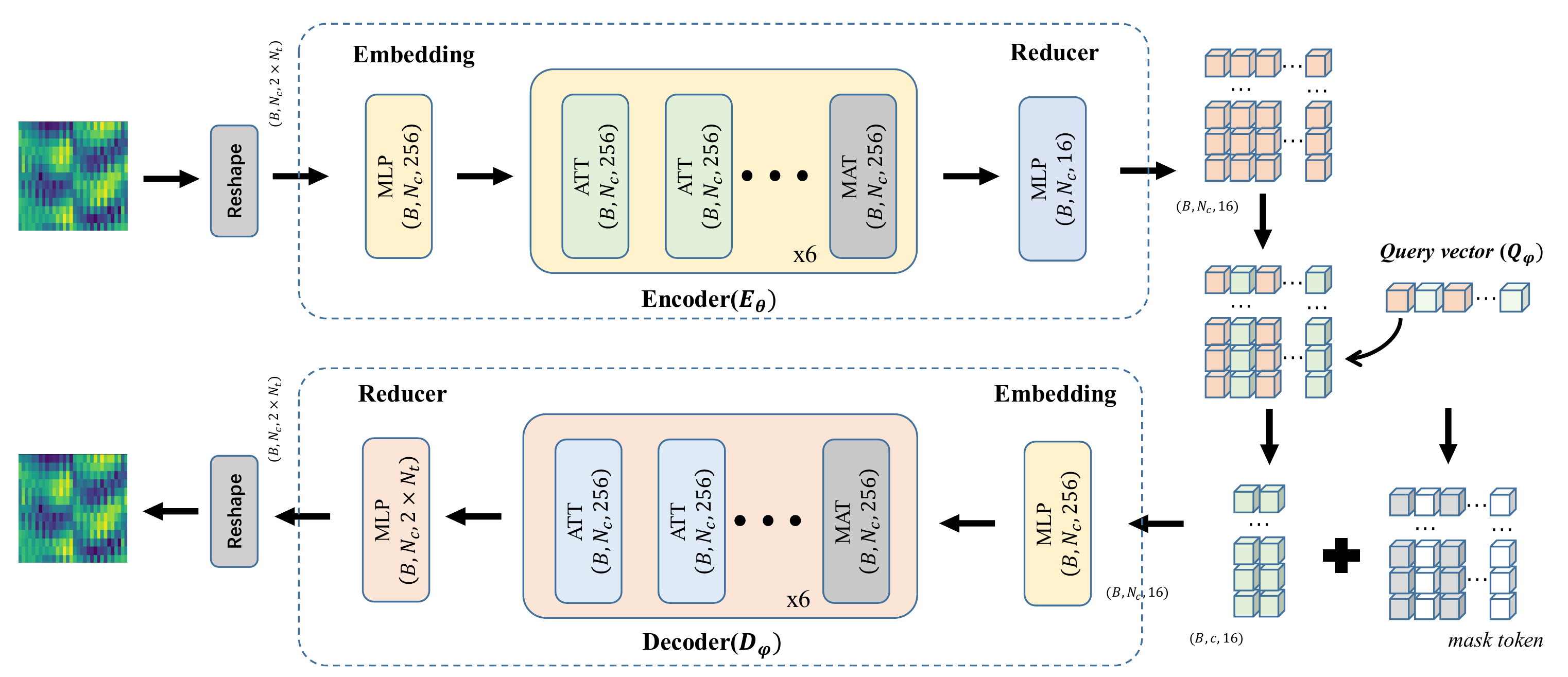}
\caption{The structure of our method. Our approach applies the typical DNN-based encoder-decoder framework and embeds the novel learnable Query vector ($Q_{\varphi}$) to active masking. First, using the based multi common attention blocks (ATT) and a mask-attention block (MAT), the latent variables are obtained. Then, we use active mask for the latent variables and get the compact representation. Finally, the decoder recovers the completed matrix, which embeds the mask token and concatenates the compact representation.}
\label{fig:Flow Mat wholenetwork}
\end{figure*}

\subsection{Joint Channel Estimation and Channel Feedback}
\subsubsection{FlowMat for Channel Estimation}
\begin{figure}[t]
\centering
\includegraphics[width=0.44\textwidth]{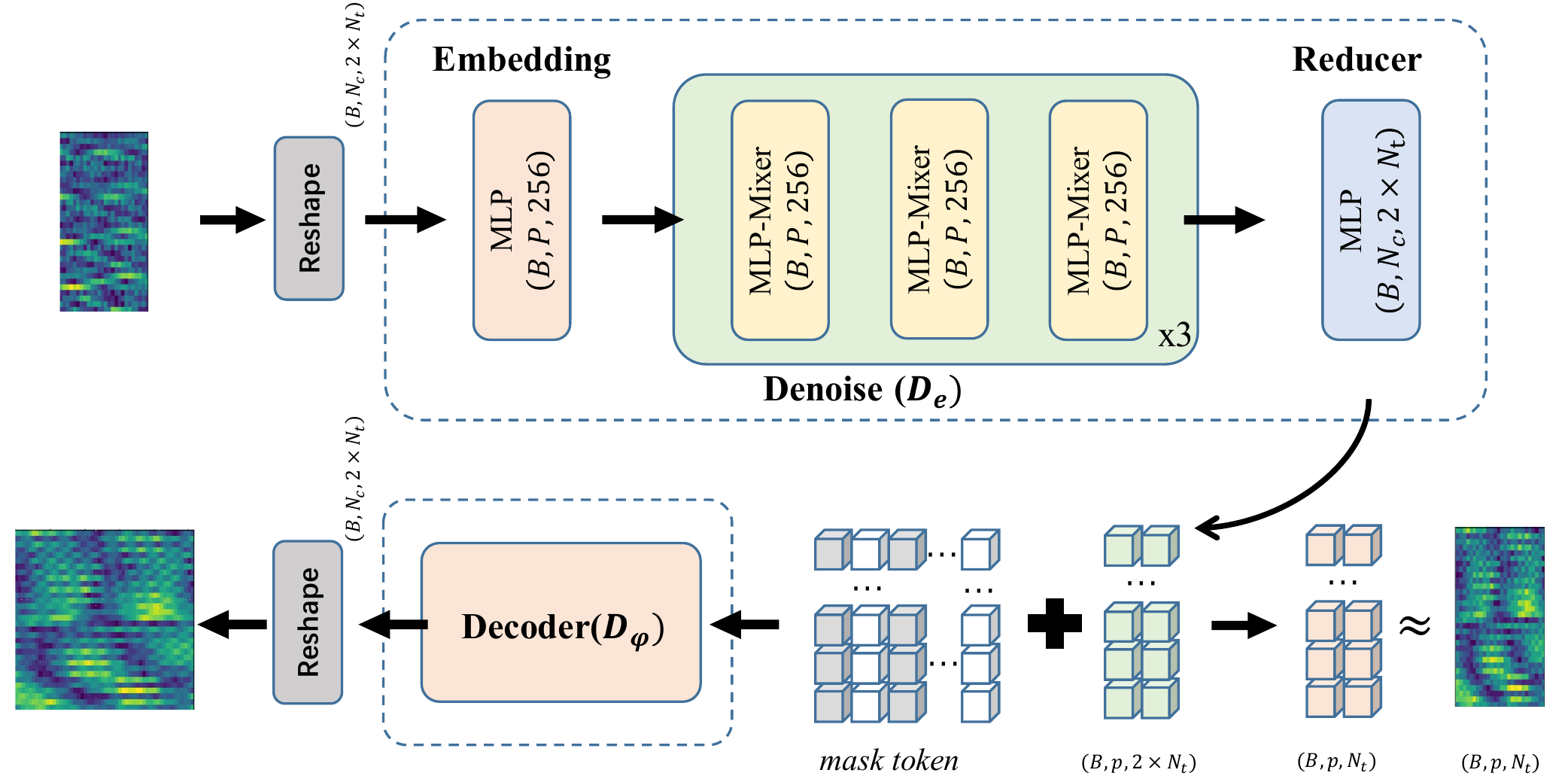}
\caption{The structure of the channel estimation network. Denoising is needed before completion. We use the lightweight MLP-Mixer to achieve encoder $D_e$, and the decoder recovers the complete channel matrix with FlowMat's decoder $D_\phi$.}
\label{fig:estimation}
\vspace{-6pt}
 \end{figure}

%Channel estimation aims to recover all $N_c$ subcarriers channel matric from the BS transmitting the $N_P$ pilots information. 
Channel estimation aims to recover all $N_c$ subcarries channel matrix from the $N_P$ pilots information. 
However, the pilots obtained by the receiver contain significant environmental noise, which further hinders the accurate estimation. 
%To recover the DL CSI with high quality, we need to build a denoise network to get pilots without noise. Then, utilizing the clean pilots recovering the whole DL CSI by the completing network.
To recover DL CSI with high quality, we need to build a denoising network to remove noise from the pilots. Then, the clean pilots can be used to recover the complete DL CSI with the help of a completion network.
As shown in Fig.~\ref{fig:estimation}, our channel estimation network includes three main components: denoising block, mask token embedding block, and recovery decoder.

For the complex values of the CSI matrix, we first concatenate the real and imaginary as inputs ($\mathbb{R}^{N_c\times{N_t}\times{2}}$). 
%The channel estimation acts at the UE side with the calculation resource constraint. Also, our design consideration is on the low complexity, then using the lightweight MLP-Mixer to achieve denosie. The network reduces the dimension of the given pilot information from the feature dimension to the original dimension, and then uses the pixel-wise loss to constrain pilot information to be the same as the noise-free full information in the ground truth.
The channel estimation in our system is performed at the user equipment (UE) side, taking into account the constraint of calculation resources. Our design focuses on achieving low complexity, and to achieve denoising, we utilize the lightweight MLP-Mixer. The network employed in our system reduces the dimension of the pilot information from the feature dimension to the original dimension. Additionally, we use a pixel-wise loss to ensure that the pilot information aligns with the noise-free full information provided in the ground truth.
After obtaining the clean pilot, we embed the masked tokens and recovery decoder as same as the FlowMat decoder ($D_\phi$). 
% network to reduce the dimension of the given pilot information from the feature dimension to the original dimension, and then use the pixel-wise loss to constrain pilot information to be the same as the noise-free full information in the ground truth.
%
% Then we embed the masked tokens on the decoder side and use the transformer to complete the decoding.
Finally, the NMSE between the output and noiseless full information is calculated.

\subsubsection{FlowMat for Channel Feedback}
The purpose of the channel compression and feedback network is to compress the wireless information in the user equipment and recover it as well as possible in the base station under the given bit bandwidth. 
In our work, we propose a transformer-based architecture in the encoder and decoder. 
As shown in Fig.~\ref{fig:Flow Mat wholenetwork}, the input is an $N_c\times{N_t}$ eigenvector matrix. Because the values of the eigenvector are complex, we will concatenate the real part and the imaginary part to obtain the $N_c\times{2}\times{N_t}$ input.
And then this input is sent to the encoder. We regard the subcarriers as the tokens of FlowMat's encoder ($E_\theta$). 
After getting the deep features $Z$, the network learns a query vector to select specific deep subcarriers' feature and only retains $m$ of them ($m$ is a super parameter), which is denoted as $Z_{Part}$.
%
%After obtaining $Z_{Part}$,
For quantization, there are two options: uniform quantization and VQ quantization\cite{nips2017_VQVAE}. 
We experimentally compare them in Sec.~\ref{sec:csi_compression}. 
The bit stream transmitted by uniform quantization is the quantized sequence of $Z_{Part}$, while VQ quantization transmits the indexes of the corresponding vectors in the codebook corresponding to $Z_{Part}$.%, which is, 

% \begin{equation}
% \mathrm{Q_{vq}}\left(z_{e,j}\right)=\min \left\|e_k-z_{e,j}\right\|_2^2.
% \end{equation}

% First, we split the tokens into several groups of sub-tokens according to channels. Each group of sub-tokens, $z^{i}_{e}$ is then processed by VQ quantization $i$. Each VQ quantization module has a learnable discrete codebook (the elements $K$ of the codebook is a preset super parameter). Unlike autoencoder, where $z^{i}_{e}$ is directly input into the decoder, $z^{i}_{e}$ in VQ quantization needs to be replaced according to the codebook, that is,
% \begin{equation}
% \mathrm{Q}\left(z_{e,j}\right)=\min \left\|e_k-z_{e,j}\right\|_2^2
% \end{equation}
% %
% It means that the network finds the codeword $e_{k}$ corresponding to $z_{e,j}$, replaces $e_{k}$ with $z_{e,j}$ one by one, and generate $z_{q}$.
% Finally, all the outputs of the VQ quantization are concatenated.
% %
% It is noted that we only need to send the index of each VQ quantization from user equipments to base stations.

Following quantization, masked tokens are utilized to recover the information, similar to the decoder process in channel estimation. Finally, the decoder is employed to recover the full information.

\subsubsection{FlowMat for Joint and Individual Tasks}
%The overall pipeline of joint channel Estimation and Channel Feedback is both shown in \textcolor{red}{Fig.?}.
%
Firstly, the channel estimation model is employed to estimate the missing channels, thereby accomplishing frequency domain interpolation and denoising.
Then, based on the estimated channel, the eigenvectors to be feedback on different subcarriers are calculated. Finally, the channel feedback model is used to implement the channel eigenvector compression and feedback.
We build a model using an end-to-end training manner without modifying the network structure to realize information estimation and feedback.
The model takes pilots as input and produces eigenvectors as output. 
We also separately train a channel estimation network and a channel feedback network, each referred to as 'splited', and subsequently evaluate their performance together.

\subsection{Training Strategy}
%Since our channel estimation network has two kinds of losses, 
The proposed network uses progressive training and joint training.
The progressive training strategy initially employs the normalized root mean square error (NMSE) loss to train both the encoder and the denoised constrained network, which is formulated as:
\begin{equation}
\label{eq:loss1_1}
L_{\mathrm{CE_{1}}}=\sqrt{\frac{\sum_{n=1}^T\sum_{k=1}^{N_{p}}\left(D_e(y_{n,k})-H_{n,k}\right)^2}{\sum_{n=1}^T \sum_{k=1}^{N_{p}} D_e(y_{n,k})^2}},
\end{equation}
where $D_e$ is the `Denoise' network in Fig.~\ref{fig:estimation}, $y_{n,k}$ is the kth channel of the input pilot, $T$ is the number of training samples, and $H^{\prime}_{i,k}$ is the selected ideal channel vector corresponding to  $y_{n,k}$.
And the decoder is trained with the NMSE loss while keeping the parameters of the encoder and the denoised constrained network fixed.
The loss function can be formulated as:
\begin{equation}
\label{eq:loss1_2}
L_{\mathrm{CE_{2}}}=\sqrt{\frac{\sum_{n=1}^T \sum_{k=1}^{N_{c}} \left(D_e(D_{\phi}(y_{n,k}))-H_{n,k}\right)^2}{\sum_{n=1}^T \sum_{k=1}^{N_{c}} D_e(D_{\phi}(y_{n,k}))^2}}
\end{equation}
where $T$ is the number of training samples, and $H_n$ is the whole ideal channel matrix.
%, and $(\cdot)^{\mathrm{H}}$ is Hadamard transpose.
%
The joint training is to train all the parts in the channel estimation network, simultaneously, which combines Eqn.~\ref{eq:loss1_1} and Eqn.~\ref{eq:loss1_2} as loss functions.

The loss function of channel compression and feedback network is,
\begin{equation}
\label{eq:score1}
L_{\mathrm{CF}} =\frac{1}{T} \sum_{j=1}^{T} \frac{1}{N_c} \sum_{i=1}^{N_{c}} \frac{\left\|w_{i, j}^{\mathrm{H}} w_{i, j}^{\prime}\right\|}{\left\|w_{i, j}\right\|\left\|w_{i, j}^{\prime}\right\|},
\end{equation}
where $N_c$ is the number of subcarriers of each sample, $w_{i, j}$ and $w_{i, j}^{\prime}$ are the eigenvectors of labels and predicted eigenvectors, respectively.

\section{Experiments}

\subsection{Data Description}
The dataset that we use is Mobile Communication Open Dataset in 2022~\footnote{This dataset is public and can be downloaded from: \url{https://www.mobileai-dataset.cn/html/default/zhongwen/shujuji/index.html?\\index=1}}.
%The data that we use in this article are from Track-1 of the third Wireless Communication AI Competition in 2022~\footnote{Mobile Communication Open dataset: \url{https://www.mobileai-dataset.cn/html/default/zhongwen/shujuji/index.html?\\index=1}}.
%
It has 32 transmit antennas and 4 receive antennas. Channel estimation and channel feature feedback are required according to specific pilots.
This dataset provides two different received pilots, namely, high-density (HD) pilot and low-density(LD) pilot as the input information. 
Each of them includes 300,000 samples. 
The high-density pilots occupy 26 resource blocks out of 52 resource blocks (odd numbers of the resource blocks) to send pilot information. 
The dimension of each data sample is $4  \times 208 \times 4$. The first `4' refers to four receiving antennas. `208' refers to 208 subcarriers (26 resource blocks, each resource block has 8 subcarriers with pilot information), and the second `4' refers to 4 OFDM symbols.
For low-density pilots, 6 of 52 resource blocks are occupied (serial numbers: 7, 15, 23, 31, 39, 47). The dimension of each data sample is $4  \times  48  \times 4$, where `48' refers to 6 resource blocks, each of which has eight subcarriers with pilots.

The time-domain full channel information contains a total of 300,000 samples as the unified information for both the high-density pilots and low-density pilots. The dimension of each sample is $4 \times 32 \times 64$, corresponding to 4 receiving antennas, 32 transmitting antennas, and 64 delay sampling.
The time-domain full channel information can convert to the eigenvector information for better performance in channel compression and feedback.
%In channel compression and feedback, the eigenvector information of the channel can be selected as input/output.
These channel eigenvectors are given by the dataset. The transmission bandwidth is divided into 13 subcarriers, each of which contains 32-dimensional eigenvectors. There are 300,000 samples in this part, which are used as unified labels for high-density and low-density pilots. The dimension of each sample is $32 \times 13$.

We use the first 95\% of the samples in the dataset for training and the last 5\% for testing, namely, the test set contains 15,000 samples.

\subsection{Comparison Results} %with the SOTA methods
We first compare our method with the SOTA methods of the two tasks individually and then conduct experiments on the joint channel estimation and feedback.
For channel estimation, we use NMSE to evaluate the error between the desired and estimated channels, which is calculated by: 
\begin{equation}
\label{eq:shown}
\mathrm{NMSE} = \frac{\sum_{n=1}^T\sum_{k=1}^{N_{p}}\left(D_e(y_{n,k})-H_{n,k}\right)^2}{T N_{p}\sum_{n=1}^T \sum_{k=1}^{N_{p}} D_e(y_{k})^2},
\end{equation}
where $y$ is the input pilots and $H^{\prime}$
is the ideal channel matrix.
To present the results clearly, we express the NMSE in decibels (dB).

For channel feedback, we employ the cosine similarity between the channel eigenvector of the feedback and the label as metric, which is calculated as:
\begin{equation}
\label{eq:score2}
\text { Rho }=\frac{1}{T} \sum_{j=1}^{T} \frac{1}{N_{c}} \sum_{i=1}^{N_{c}} \frac{\left\|w_{i, j}^{\mathrm{H}} w_{i, j}^{\prime}\right\|}{\left\|w_{i, j}\right\|\left\|w_{i, j}^{\prime}\right\|},
\end{equation}
where $T$ is the number of tested samples,  $N_{c}$ is the number of subcarriers of each sample, $w_{i, j}$ and $w_{i, j}^{\prime}$ are the eigenvectors of labels and predicted eigenvectors, respectively.

\subsubsection{Channel estimation}
We compare ours with four SOTA channel estimation methods, ChannelNet~\cite{MehranSoltani2019DeepLC}, AttentionNet~\cite{DianxinLuan2022AttentionBN},
CBDNet \cite{YuJin2020ChannelEF}, and Attention-Aid \cite{JiabaoGao2021AnAD}.
ChannelNet regards the problem as the image super-resolution and image denoising. 
In ChannelNet, the pilots are regarded as low-resolution images, and the super-resolution network, SRCNN~\cite{ChaoDong2014ImageSU} cascaded with the denoising network, DNCNN~\cite{KaiZhang2016BeyondAG} is used to estimate the channels.
AttentionNet proposes a new hybrid encoder-decoder structure, It contains a transformer encoder and a residual decoder CNN.
An improved CNN network is proposed in CBDNet, which uses noise level estimation subnetwork, non-blind denoising sub-network and asymmetric joint loss function for channel estimation.
Attention-Aid introduces a new attention-aided deep learning channel estimation framework for traditional large-scale MIMO systems, which includes channel attention modules.

We use Eqn.~\ref{eq:score2} for evaluation and the performance is shown in Tab.~\ref{tab:estimationonly}.
It shows that our method is better than other models in the channel estimation only task.
Ours achieves -6.6106 and -5.9438 in the high-density test set and low-density test set, respectively (lower is better), which improves CBDNet by about 14\% and 11\%.
When compared with the attention-based method, Attention-Aid, our method improves Attention-Aid by 21\% and 39\%.
Our masked token scheme has learned a better mapping between deep features and channel information by
adding learnable tokens, which achieves better estimation results.
ChannelNet is a CNN-based method that performs denoising and estimation in two separate stages. Although their method achieves excellent results on high-density datasets (ranked second among all methods), its performance on low-density datasets is not satisfactory (ranked last among all methods). In contrast, our method can achieve good performance in both settings.
For computation complexity, the Floating Point Operations (denoted as `FLOPs') and network parameters (denoted as `Params') results are shown in \ref{tab:Ablationcomplex_1}. 
Among the comparison methods, FlowMat exhibits the lowest FLOPs with slightly higher Params.

\begin{table}[t]
  \centering\small
  \caption{Quantitative comparison of channel estimation only on the dataset. Lower is better.}
  \vspace{-1mm}
  %The best results are in \textbf{bold}.}
  %\resizebox{0.8\linewidth}{!}{
  \setlength{\tabcolsep}{4mm}{
  \renewcommand{\arraystretch}{1.1}  
  \begin{threeparttable}
  \begin{tabular}{c|c|c}
    \whline
      Method & HD NMSE (dB) & LD NMSE (dB)\\
     \hline
      ChannelNet&  -6.3720 & -2.2810\\
      CBDNet&  -5.6891  &  -5.2179\\
      AttentionNet&  -5.3929&-2.8207\\
      Attention-Aid&  -5.1555	&	-4.0300\\
      \hdashline
      Ours&  \textbf{-6.6106}&\textbf{-5.9438}\\
    \whline
  \end{tabular}
  \end{threeparttable}}
  \vspace{-1mm}
  \label{tab:estimationonly}
\end{table}

% \begin{comment}
\begin{table}[t]
  \centering \small
  \caption{Quantitative comparison of channel compression and feedback only on the dataset.}
  \vspace{-1mm}
  %The best results are in \textbf{bold}.}
  %\resizebox{0.75\linewidth}{!}{
  \setlength{\tabcolsep}{5mm}{
  \renewcommand{\arraystretch}{1.1}
  \begin{threeparttable}
  \begin{tabular}{c|c|c|c}
    \whline
      Method & 64 bit & 128 bit & 256 bit\\
    \hline
      CsiNet&  0.7717 & 0.8140 & 0.8433 \\
      CsiNet+&  0.7634& 0.7685 & 0.7894\\
      ImCsiNet& 0.8514 & 0.8825 & 0.9240\\
      DCRNet &0.7764 & 0.7965 & 0.8611\\
      EVCsiNet &0.8308 & 0.8687 & 0.9247\\
      EVCsiNet-T &0.8532&0.9292&0.9632\\
      \hdashline      
      Ours(VQ) & 0.8922 & 0.9283  & 0.9464 \\
      Ours(UQ) & \textbf{0.9015} &  \textbf{0.9333} & \textbf{0.9648} \\
    \whline
  \end{tabular}
  \end{threeparttable}}
  \vspace{-6pt}
  \label{tab:compressiononly}
\end{table}
% \end{comment}

%  \begin{figure}[t]
% \centering
% \includegraphics[width=0.48\textwidth]{pic/table2.pdf}
% \caption{Quantitative comparison of channel compression and feedback only on the dataset. Higher is better.}
% \label{fig:table2}
%  \end{figure}

 \begin{table}[t]
\centering 
\caption{Comparison of computational complexity of estimation methods. In the 1st line (Setting) `High density' and `Low density' mean high and low density pilot dataset respectively.}

\setlength{\tabcolsep}{6mm}{
\renewcommand{\arraystretch}{1.1}
\begin{threeparttable}
\begin{tabular}{c|c|c}
\whline %\toprule
Setting &\multicolumn{1}{c|}{High density}&\multicolumn{1}{c}{Low density}\\
\hline
Method & FLOPs/Params & FLOPs/Params \\
\hline  %\midrule
ChannelNet	&2.59G/0.194M &2.59G/0.194M\\
Attention-Aid	& 2.25G/0.169M& 2.25G/0.169M\\
AttentionNet	& 3.95G/5.29M & 9.12G/5.28M \\
CBDNet &	7.76G/4.33M&	7.76G/4.33M\\
OursEstimation	&1.18G/5.30M &1.04G/5.30M\\
\whline %\bottomrule   
\end{tabular}
\end{threeparttable}}
\vspace{-6pt}
\label{tab:Ablationcomplex_1}
\end{table}

 \begin{table}[t]
\centering 
\caption{Comparison of computational complexity of feedback methods. `n Bit' means the compression bit.}
  %low 6 ,high 26
%\resizebox{16.2cm}{!} {
%\resizebox{0.75\linewidth}{!}{
\setlength{\tabcolsep}{6mm}{
\renewcommand{\arraystretch}{1.1}
\begin{threeparttable}
\begin{tabular}{c|c|c}
\whline %\toprule
Setting &\multicolumn{1}{c|}{64 Bit}&\multicolumn{1}{c}{128 Bit}\\
\hline
Method & FLOPs/Params & FLOPs/Params \\
\hline  %\midrule
		
CsiNet&	1.63M/0.217M&	 1.87M/0.430M\\
CsiNet+&	9.51M/0.236M & 9.78M/0.449M\\
DCRNet	&0.99M/0.215M & 1.36M/0.429M \\
EVCsiNet&	1.60G/5.18M & 1.61G/5.65M\\
EVCsiNet\_T&	0.21G/16.67M& 0.21G/17.52M\\
ImCsiNet&	0.83G/821M&	 0.84G/	821M\\
OursCompress	&1.53G/	9.51M  & 1.54G/9.55M\\
\hline %\midrule
JCEF&	2.54G/1.47M & 2.54G/2.75M \\
\whline %\bottomrule   
\end{tabular}
\end{threeparttable}}
\vspace{-6pt}
\label{tab:Ablationcomplex_2}
\end{table}

\subsubsection{CSI compression and feedback only}\label{sec:csi_compression}
For CSI compression and feedback, we also compare with six SOTA methods, CsiNet~\cite{ChaoKaiWen2018CsiNet}, CsiNet+~\cite{GuoJ2020CsiNet+}, ImCsiNet~\cite{MuhanChen2021DeepLI}, DCRNet~\cite{ShunpuTang2022DilatedCB}, EVCsiNet~\cite{WendongLiu2021EVCsiNetEC}, and EVCsiNet-T~\cite{HanXiao2021AIEW}.
CsiNet and CsiNet+ are deep-learning-based feedback neural networks, which contain Convolution, fully-connected layers and BatchNorm layers.
ImCsiNet is an implicit feedback neural network, which contains LSTM structure, fully-connected layers and BatchNorm layers.
DCRNet is a CSI feedback network based on dilated convolution. The dilated convolution layer is used to enhance the reception field without increasing the convolution size.
EVCsiNet is an eigenvector based deep learning CSI feedback method, in which the joint eigenvector vectors cascaded from multiple subcarriers are compressed and restored at the encoder and decoder respectively. 
The network of EVCsiNet adopts the architecture of Residual block~\cite{KaimingHe2015DeepRL}.
EVCsiNet-T has developed the method of EVCsiNet, including i) channel data analysis and pre-processing, ii) neural network design and iii) quantization enhancement.

We use the Eqn.~\ref{eq:score2} for evaluation and the performance is shown in Tab.~\ref{tab:compressiononly}.
%Tab.~\ref{tab:compressiononly} presents the comparison results in terms of correlation. 
We report the performance of compressing and recovering using original eigenvector vectors. 
Tab.~\ref{tab:compressiononly} also shows that our way is better than other models in all bits overhead. 
For example, it achieves 0.9015 in the 64-bit, which improves CsiNet by 14.53\%. 
FlowMat and EVCsiNet-T use the same transformer structure,  and FlowMat improves the latter by 3.59\% in 64 Bit.
The main reason is that the last layer of the encoder in EVCsiNet-T directly decreases its output into low dimension vector. 
This approach results in significant information loss, causing difficulties for the decoder in recovering the original data.
The eigenvector is denser than DFT, thus the CNN ways' performance is lower than that of LSTM and Transformer ways, which utilizes the correlation in subcarriers. 
In CNN ways, the CsiNet+ uses the 7*7 kernel to obtain the key feature, which lost more information than CsiNet's 3*3 kernel in eigenvector data. 
Similar to channel estimation only to estimate computation complexity, the experimental results w.r.t. FLOPs and Params are shown in \ref{tab:Ablationcomplex_2}. 
The FLOPs of ours is lower than EVCsiNet and CsiNet and the Params is lower than EVCsiNet\_T and ImCsiNet.

\begin{comment}
\begin{table}[t]
  \centering\small
  \caption{Quantitative comparison of channel compression and feedback only on the dataset.}
  \vspace{-1mm}
  %The best results are in \textbf{bold}.}
  \resizebox{8.4cm}{!} {
  \begin{tabular}{cc|ccc}
    \toprule
      &Method & 64 bit & 128 bit & 256 bit\\
      \midrule
      \multirow{4}{*}{High density} & JCEF&0.7229 &0.7970 &0.8558\\
      & AttentionNet+EVCsiNet-T&0.8329 & 0.8965 &0.9257\\
      & AttentionNet+DCRNet&0.7599 &0.7786 & 0.8361\\

      &Ours& 0.8556 & 0.9188 & 0.9406 \\
       \midrule
      \multirow{4}{*}{Low density} & JCEF&0.7031 & 0.7590 &0.8335\\
      & AttentionNet+EVCsiNet-T&0.8061 &0.8410 &0.8568\\
      & AttentionNet+DCRNet&0.7411 &0.7549 & 0.8058\\

      &Ours& 0.8439 & 0.8936 & 0.9098 \\
    \bottomrule
    
  \end{tabular}}
  \vspace{-2mm}
  \label{tab:joint}
\end{table}
\end{comment}

 % \begin{figure}[t]
 % 		\centering
 % 		\includegraphics[width=0.8\linewidth]{pic/table3.pdf}
 % 		\caption{Quantitative comparison of joint channel estimation and channel compression on the dataset. Higher is better.}
 % 		\label{fig:table3}
 % \end{figure}

\begin{figure*}[t]
   \centering   
   \subfigure[High Density]{\includegraphics[width=0.48\linewidth,height=0.33\linewidth]{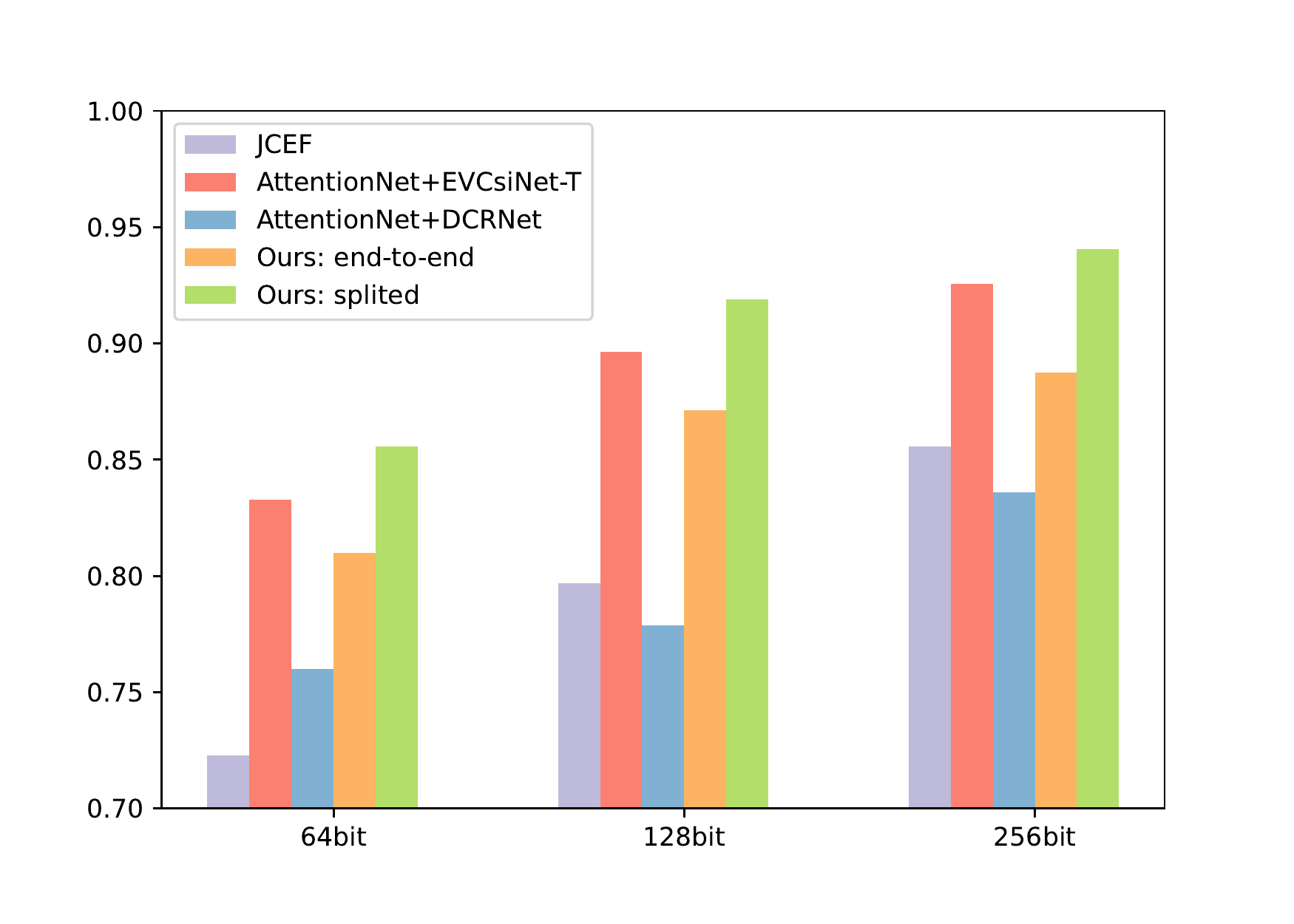}\label{fig:High}}
   \subfigure[Low Density]{\includegraphics[width=0.48\linewidth,height=0.33\linewidth]{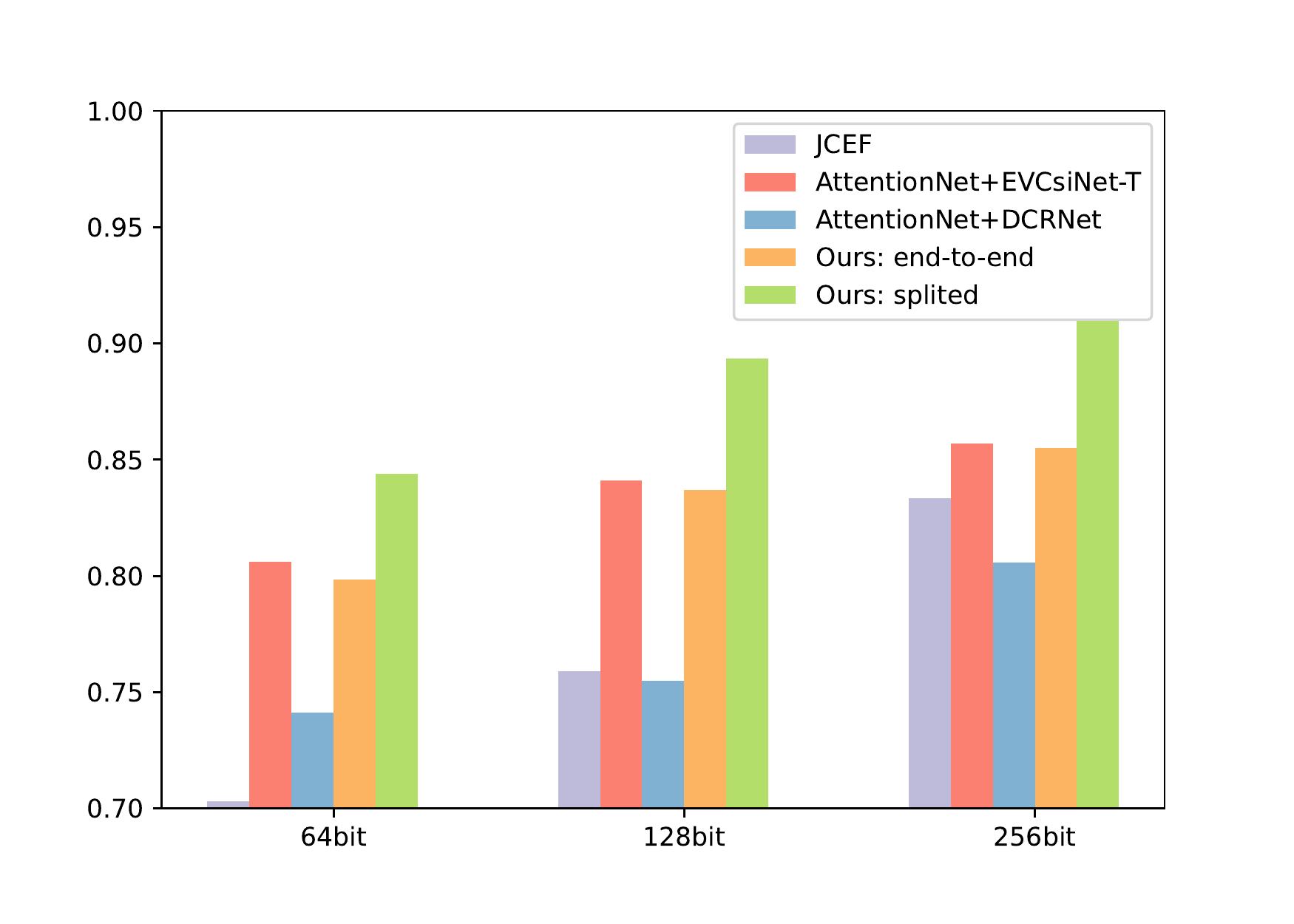}\label{fig:Low}}
\caption{Quantitative comparison of joint channel estimation and channel compression on the dataset. Higher is better.}
\vspace{-6pt}
\label{fig:E2E_Desity_Performance}
\end{figure*}

\subsubsection{Joint Channel Estimation and Feedback}
Since there are few joint channel estimation and feedback methods, we only compare one joint channel and feedback method, namely JCEF~\cite{ChenTong2020DeepLF}.
JCEF which completes the full channel information, compresses the information using uniform quantization, and restores the information at last.  
Meanwhile, we combine some of the channel estimation methods with the channel feedback method to compare with our method.
It denotes `AttentionNet+EVCsiNet-T' and `AttentionNet+DCRNet'.

We use Eqn.~\ref{eq:score2} for evaluation and the performance is shown in Fig.~\ref{fig:High}. 
In low-density pilots (Fig.~\ref{fig:Low}), we can see that the Rho increases by 0.0378, 0.0526, and 0.053 on 64-bit, 128-bit and 256-bit respectively when using our model. 
In high-density pilots (Fig.~\ref{fig:High}), we can see that the Rho increases by 0.0227, 0.0223, and 0.0149 on 64-bit, 128-bit and 256-bit respectively when using our model. 
These results show the effectiveness of our FlowMat in joint channel estimation and channel compression.

\subsection{Ablation Study}
In this subsection, we do the ablation study using the low-density dataset.

% \begin{table}[t]
%   \centering\small
%   \caption{Ablation study: End-to-end solution V.S. splited training in Low-density Pilot.}
%   %The best results are in \textbf{bold}.}
%   \resizebox{7.6cm}{!} {
%   \renewcommand{\arraystretch}{1.1}
%   \begin{tabular}{c|c|ccc}
%     \whline
%       Method & Pilot density & 64 bit & 128 bit & 256 bit\\
%       \hline
%       \multirow{2}*{End2end} & Low & 0.7986 & 0.8370 & 0.8549\\
%       ~ & High & 0.8099 & 0.8712 & 0.8875\\
%       \hline
%       \multirow{2}*{Ours} & Low    & 0.8439 & 0.8936 & 0.9098  \\
%       ~&  High   & 0.8556 & 0.9188 & 0.9406  \\
%     \whline  
%   \end{tabular}}
%   \label{tab:Ablationsolution}
% \end{table}

\begin{table}[t]
  \centering\small
  \caption{Ablation study: The loss function for channel estimation.}
  \label{tab:Ablationloss}
  %The best results are in \textbf{bold}.}
  %\resizebox{7.8cm}{!}{
  %\resizebox{0.8\linewidth}{!}{%
  \setlength{\tabcolsep}{3mm}{
  \renewcommand{\arraystretch}{1.2}
  \begin{tabular}{c|ccc}
    \whline
      Method  & L1+L1 & L1+NMSE & NMSE+NMSE \\
      \hline
      High density & 0.9748 & 0.9770 & 0.9774\\
      Low density& 0.9330 & 0.9363  & 0.9366\\
    \whline
  \end{tabular}}
  \vspace{-12pt}
\end{table}

\begin{comment}
\begin{table}[t]
  \centering\small
  \caption{Ablation study: The Way of token reduction in CSI compression.}
  %The best results are in \textbf{bold}.}
  \resizebox{7.8cm}{!} {
  \begin{tabular}{c|ccc}
    \toprule
      Method & MLP & Token merging & Ours \\
      \midrule
      64 bit & 0.9026 & 0.9018 & 0.9015 \\
      128 bit & 0.9295 & 0.9339 & 0.9333 \\
      256 bit & 0.9634 & 0.9588 & 0.9648 \\
    \bottomrule
    
  \end{tabular}}
  \label{tab:AblationWay}
\end{table}
\end{comment}

 \begin{figure}[t]
 \centering
\includegraphics[width=0.8\linewidth,height=0.6\linewidth]{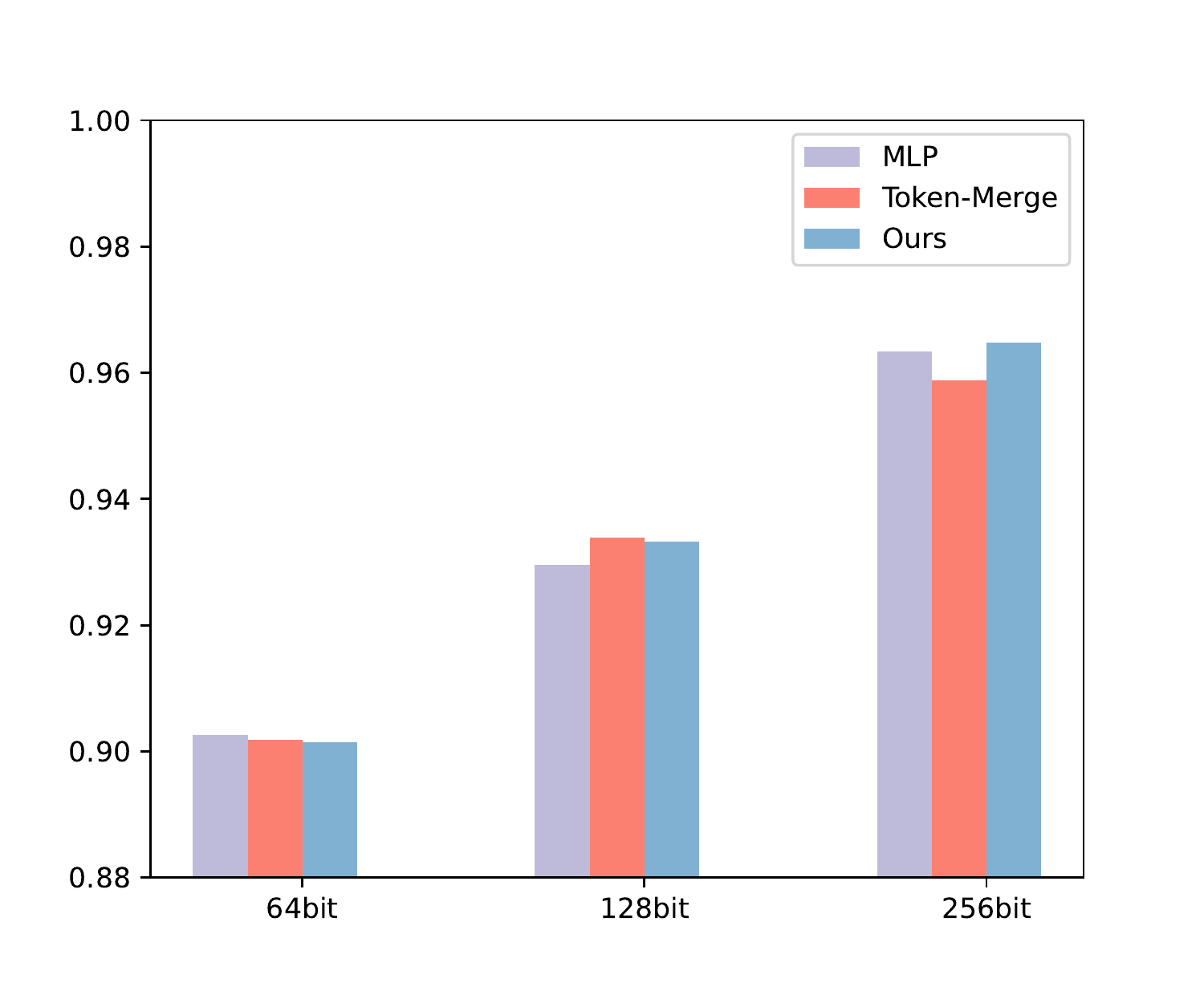}
 \vspace{-18pt}
 \caption{Ablation study: The way of token reduction in CSI compression.}
 \label{fig:table6}
\vspace{-12pt}
 \end{figure}
 
\subsubsection{End-to-end solution v.s. splited training}

\begin{table}[t]
  \centering
  \caption{Ablation study: Tokens' content and whether updating. U represents parameter update. The best and second best results are shown in \textbf{bold} and \underline{underlined}.}
  %The best results are in \textbf{bold}.}
  \resizebox{\linewidth}{!}{
  %\resizebox{13.8cm}{!} {
  % {
  \begin{tabular}{c|cccc}
    \whline 
      Method & zero+ U & randn+ U & zero+ w.o. U & randn+ w.o. U \\
      \hline
      64 bit  & 0.9015 & \bf{0.9036} &0.9023 & \underline{0.9030}   \\
      128 bit & \underline{0.9333} & 0.9315 & 0.9330 & \bf{0.9348}\\
      256 bit & \bf{0.9648} & \underline{0.9635}  & 0.9618 & 0.9588\\
    \whline    
  \end{tabular}}
  \label{tab:AblationTokensstudy}
  \vspace{-6pt}
\end{table}

To show the effectiveness of the splited training pipeline, we build a model named
`Ours: splited' in Fig~\ref{fig:High}, and Fig~\ref{fig:Low} is tested on high-density pilots. 
We compared it with the end-to-end training manner to realize information estimation without modifying the network structure.
The input is feedback with pilots and the output is eigenvectors. 
And Eqn.~\ref{eq:score1} serves as the loss function. 
In low-density pilots, it can be observed that Rho of splited manner is higher than end-to-end on 64-bit, 128-bit, and 256-bit, respectively.
The same phenomenon can be observed in high density.
And our method in splited manner performs best among the competing methods.
When compared to other end-to-end methods, our approach still outperforms them (e.g., see JCEF and AttentionNet+DCRNet), making it the most effective in the end-to-end framework.
%In high-density pilots, we can see that the Rho is lower than Individual model on 64-bit, 128-bit, and 256-bitt respectively when using Joint model.
%
%The results show the effectiveness of our pipeline.

\subsubsection{The loss function for channel estimation}
For channel estimation, we also try other loss functions for the ablation study. In the original model, NMSE loss and NMSE loss are used for channel denoising channel estimation, respectively. And In Table~\ref{tab:Ablationloss}, it is denoted as `NMSE+NMSE'. And we build some models for comparison, `L1+L1' and  `L1+NMSE'. 
The one before the `+' refers to the loss used for denoising, and the one after the `+' refers to the loss used for estimation.
We can see that among these methods, `NMSE+NMSE' performs the best result.
For example, it exceeds `L1+L1' and `L1+NMSE' by about 0.0036 and 0.0003 respectively in the low-density test set.

\subsubsection{The way of token reduction in CSI feedback}

In our channel compression and feedback model, we propose channel selection technique and mask embedding for token reduction and expansion, respectively.
There are also some other methods that can process token reduction and expansion.
For example, MLP layers are a common way for dimension reduction and expansion~\cite{ChenTong2020DeepLF,DianxinLuan2022AttentionBN}.
We use the MLP layers and token average merging for comparison.
For MLP layers, we use one MLP for token reduction and another MLP for token expansion.
For token average merging, we group all the input tokens equally and then calculate the average value of each group for token reduction.
For token expansion, the way is the same as the original model.
We do the experiments on channel compression and feedback only. And the results show the effectiveness of our proposed method.
In Fig.~\ref{fig:table6}, `MLP' and `Token merging' mean using MLP layers and  token merging respectively.
It shows that although the MLP method has advantage (0.0011) over our original method in the case of low-bit compression, the performance of our method exceeds the MLP method in the middle- and high-bit compression (exceed by 0.0038 and 0.0014).
When compared with `Token merging' and ours,  our method performs equally well as the token merging method at lower bit rates, but exhibits significant advantages (exceeding by 0.006) at higher bit rates.

\subsubsection{Tokens' content and update}
The contents of masked tokens and whether updating will also affect the performance. Therefore, we try to make the initial values of masked tokens all zero (`0'), the initial values are standard normal distribution (`randn'), and the token values are updated / not updated in the training process. Therefore, there are four experimental combinations: `zero+update', `randn+update', `zero+without update', and `randn+without update'.
The results are shown in Table~\ref{tab:AblationTokensstudy}.
It can be seen that in the cases of high-bit compression and low-bit compression, random initialization and updating have different effects on performance.
When the compression bit is low (64-bit), the impact of the initialization of the standard normal distribution (`randn') is dominant, and updating has a weak impact on the results.
When the compression bit is high (128-bit), the impact of updating is dominant. Updating the tokens' parameters will boost the performance.

\subsubsection{Which tokens should be masked?}
Finally, we do the ablation study about which tokens should be masked in the mask token embedding operation. In our original model, we learn a query vector to determine which tokens should be masked. Without losing generality, we build a comparable model whose query vectors are randomly selected at the initial time and are fixed during training and testing.
The performance of learnable query vector and fixed query vector is almost the same. 
However, because the performance of the learnable query vector is slightly higher in some specific cases, we employed the technique of query learnable vector in our paper.

\section{Conclusions}
In this paper, we leverage the correlation in the inherent frequency domain among channel matrices and thereby propose a unified framework named FlowMat for joint channel estimation and feedback in massive MIMO systems. 
The proposed FlowMat takes advantage of the encoder-decoder architecture.
The entire encoder-decoder network is dedicated to channel feedback utilizing a learnable mask token, which has achieved excellent channel feedback performance. 
The decoder is used for channel estimation, 
Channel estimation is achieved by a decoder with the same structure as in channel feedback, wherein a lightweight multilayer perceptron denoising module is utilized for further accurate estimation.
We have conducted extensive, which demonstrates the superior performance on channel estimation and feedback in both joint and separate tasks. 
In addition, the FlowMat offers a significant reduction in computation complexity in particular channel estimation. 
%To improve our approach, future work includes finding a faster and more effective compression method and comparing it with more state-of-the-art methods.

{\appendices
}

\bibliographystyle{IEEEtran}
\bibliography{egbib}

\vfill

\end{document}